\documentstyle[12pt]{article}
\begin{document}

\catcode`@=11
\long\def\@caption#1[#2]#3{\par\addcontentsline{\csname
  ext@#1\endcsname}{#1}{\protect\numberline{\csname
  the#1\endcsname}{\ignorespaces #2}}\begingroup
    \small
    \@parboxrestore
    \@makecaption{\csname fnum@#1\endcsname}{\ignorespaces #3}\par
  \endgroup}
\catcode`@=12
\newcommand{\newc}{\newcommand}
\newc{\gsim}{\lower.7ex\hbox{$\;\stackrel{\textstyle>}{\sim}\;$}}
\newc{\lsim}{\lower.7ex\hbox{$\;\stackrel{\textstyle<}{\sim}\;$}}
\newc{\gev}{\,{\rm GeV}}
\newc{\mev}{\,{\rm MeV}}
\newc{\ev}{\,{\rm eV}}
\newc{\kev}{\,{\rm keV}}
\newc{\tev}{\,{\rm TeV}}
\def\ln{\mathop{\rm ln}}
\def\tr{\mathop{\rm tr}}
\def\Tr{\mathop{\rm Tr}}
\def\Im{\mathop{\rm Im}}
\def\Re{\mathop{\rm Re}}
\def\bR{\mathop{\bf R}}
\def\bC{\mathop{\bf C}}
\def\lie{\mathop{\hbox{\it\$}}} 
\newc{\sw}{s_W}
\newc{\cw}{c_W}
\newc{\swsq}{s^2_W}
\newc{\cwsq}{c^2_W}
\newc{\mgrav}{m_{3/2}}
\newc{\mz}{M_Z}
\newc{\mpl}{M_*}
\newc{\mw}{m_{\rm weak}}
%
%
\def\NPB#1#2#3{Nucl. Phys. {\bf B#1} (19#2) #3}
\def\PLB#1#2#3{Phys. Lett. {\bf B#1} (19#2) #3}
\def\PLBold#1#2#3{Phys. Lett. {\bf#1B} (19#2) #3}
\def\PRD#1#2#3{Phys. Rev. {\bf D#1} (19#2) #3}
\def\PRL#1#2#3{Phys. Rev. Lett. {\bf#1} (19#2) #3}
\def\PRT#1#2#3{Phys. Rep. {\bf#1} (19#2) #3}
\def\ARAA#1#2#3{Ann. Rev. Astron. Astrophys. {\bf#1} (19#2) #3}
\def\ARNP#1#2#3{Ann. Rev. Nucl. Part. Sci. {\bf#1} (19#2) #3}
\def\MPL#1#2#3{Mod. Phys. Lett. {\bf #1} (19#2) #3}
\def\ZPC#1#2#3{Zeit. f\"ur Physik {\bf C#1} (19#2) #3}
\def\APJ#1#2#3{Ap. J. {\bf #1} (19#2) #3}
\def\AP#1#2#3{{Ann. Phys. } {\bf #1} (19#2) #3}
\def\RMP#1#2#3{{Rev. Mod. Phys. } {\bf #1} (19#2) #3}
\def\CMP#1#2#3{{Comm. Math. Phys. } {\bf #1} (19#2) #3}
\relax
%
%
%
\def\beq{\begin{equation}}
\def\eeq{\end{equation}}
\def\bea{\begin{eqnarray}}
\def\eea{\end{eqnarray}}
%
%
%
\def\boxeqn#1{\vcenter{\vbox{\hrule\hbox{\vrule\kern3pt\vbox{\kern3pt
\hbox{${\displaystyle #1}$}\kern3pt}\kern3pt\vrule}\hrule}}}
%
%
\def\qed#1#2{\vcenter{\hrule \hbox{\vrule height#2in
\kern#1in \vrule} \hrule}}
\def\half{{\textstyle{1\over2}}} 
%
%
%
%
\newc{\ie}{{\it i.e.}}          \newc{\etal}{{\it et al.}}
\newc{\eg}{{\it e.g.}}          \newc{\etc}{{\it etc.}}
\newc{\cf}{{\it c.f.}}
%
%
%
%
\def\CAG{{\cal A/\cal G}}
\def\CA{{\cal A}} \def\CB{{\cal B}} \def\CC{{\cal C}} \def\CD{{\cal D}}
\def\CE{{\cal E}} \def\CF{{\cal F}} \def\CG{{\cal G}} \def\CH{{\cal H}}
\def\CI{{\cal I}} \def\CJ{{\cal J}} \def\CK{{\cal K}} \def\CL{{\cal L}}
\def\CM{{\cal M}} \def\CN{{\cal N}} \def\CO{{\cal O}} \def\CP{{\cal P}}
\def\CQ{{\cal Q}} \def\CR{{\cal R}} \def\CS{{\cal S}} \def\CT{{\cal T}}
\def\CU{{\cal U}} \def\CV{{\cal V}} \def\CW{{\cal W}} \def\CX{{\cal X}}
\def\CY{{\cal Y}} \def\CZ{{\cal Z}}
%
%
%
%
%
\def\grad#1{\,\nabla\!_{{#1}}\,}
\def\gradgrad#1#2{\,\nabla\!_{{#1}}\nabla\!_{{#2}}\,}
\def\partder#1#2{{\partial #1\over\partial #2}}
\def\secder#1#2#3{{\partial^2 #1\over\partial #2 \partial #3}}
%
%
%
%
%
\def\ltap{\ \raise.3ex\hbox{$<$\kern-.75em\lower1ex\hbox{$\sim$}}\ }
\def\gtap{\ \raise.3ex\hbox{$>$\kern-.75em\lower1ex\hbox{$\sim$}}\ }
\def\gl{\ \raise.5ex\hbox{$>$}\kern-.8em\lower.5ex\hbox{$<$}\ }
\def\roughly#1{\raise.3ex\hbox{$#1$\kern-.75em\lower1ex\hbox{$\sim$}}}
%
%
%
%
\def\slash#1{\rlap{$#1$}/} 
\def\dsl{\,\raise.15ex\hbox{/}\mkern-13.5mu D} 
\def\delsl{\raise.15ex\hbox{/}\kern-.57em\partial}
\def\Ksl{\hbox{/\kern-.6000em\rm K}}
\def\Asl{\hbox{/\kern-.6500em \rm A}}
\def\Dsl{\hbox{/\kern-.6000em\rm D}} 
\def\Qsl{\hbox{/\kern-.6000em\rm Q}}
\def\gradsl{\hbox{/\kern-.6500em$\nabla$}}
%
%
\let\al=\alpha
\let\be=\beta
\let\ga=\gamma
\let\Ga=\Gamma
\let\de=\delta
\let\De=\Delta
\let\ep=\varepsilon
\let\ze=\zeta
\let\ka=\kappa
\let\la=\lambda
\let\La=\Lambda
\let\del=\nabla
\let\si=\sigma
\let\Si=\Sigma
\let\th=\theta
\let\Up=\Upsilon
\let\om=\omega
\let\Om=\Omega
\def\ph{\varphi}
%
%
%
\newdimen\pmboffset
\pmboffset 0.022em
\def\oldpmb#1{\setbox0=\hbox{#1}%
 \copy0\kern-\wd0
 \kern\pmboffset\raise 1.732\pmboffset\copy0\kern-\wd0
 \kern\pmboffset\box0}
\def\pmb#1{\mathchoice{\oldpmb{$\displaystyle#1$}}{\oldpmb{$\textstyle#1$}}
	{\oldpmb{$\scriptstyle#1$}}{\oldpmb{$\scriptscriptstyle#1$}}}
%
%
%
%
%
\def\bar#1{\overline{#1}}
\def\vev#1{\left\langle #1 \right\rangle}
\def\bra#1{\left\langle #1\right|}
\def\ket#1{\left| #1\right\rangle}
\def\abs#1{\left| #1\right|}
\def\vector#1{{\vec{#1}}}
\def\inv{^{\raise.15ex\hbox{${\scriptscriptstyle -}$}\kern-.05em 1}}
\def\pr#1{#1^\prime}  
\def\lbar{{\lower.35ex\hbox{$\mathchar'26$}\mkern-10mu\lambda}} 
\def\e#1{{\rm e}^{^{\textstyle#1}}}
\def\ee#1{\times 10^{#1} }
\def\om#1#2{\omega^{#1}{}_{#2}}
\def\imp{~\Rightarrow}
\def\coker{\mathop{\rm coker}}
\let\p=\partial
\let\<=\langle
\let\>=\rangle
\let\ad=\dagger
\let\txt=\textstyle
\let\h=\hbox
\let\+=\uparrow
\let\-=\downarrow
\def\dot{\!\cdot\!}
\def\vfilll{\vskip 0pt plus 1filll}

\def\thefootnote{\fnsymbol{footnote}}
%

\begin{titlepage}
\begin{flushright}
{IASSNS-HEP-98/007\\
hep-ph/9802358\\
February 1998\\
}
\end{flushright}
\vskip 2cm
\begin{center}
{\Large\bf Supersymmetric D-term Inflation, Reheating and
Affleck-Dine Baryogenesis\footnote{Research
supported in part by U.S.\ Department of Energy contract
\#DE-FG02-90ER40542, by the W.M.~Keck Foundation, and through the
generosity of Helen and Martin Chooljian. \\
Email: {\tt kolda@sns.ias.edu, jmr@sns.ias.edu.}}}
\vskip 1cm
{\large
Christopher Kolda
and John March-Russell\footnote{Alfred P.~Sloan Foundation Fellow.}}\\
\vskip 0.5cm
{School of Natural Sciences\\
Institute for Advanced Study\\
Princeton, NJ~08540\\}
\end{center}
\vskip .5cm
\begin{abstract}
The phenomenology of supersymmetric models of inflation, where the
inflationary vacuum energy is dominated by $D$-terms of a U(1), is
investigated.  Particular attention
is paid to the questions of how to arrange for sufficient e-folds
of inflation to occur, what kind of thermal history is expected after
the end of inflation, and how to implement successful baryogenesis.
Such models are argued to require a more restrictive
symmetry structure than previously thought. 
In particular, it is non-trivial that the decays of the fields 
driving $D$-inflation can reheat the universe in such a way as to
avoid the strong gravitino production constraints. We also show
how the initial conditions for Affleck-Dine baryogenesis can arise in
these models and that the simplest flat directions along which baryon
number is generated can often be ruled out by the constraints coming from
decoherence of the condensate in a hot environment. At the end, we find
that successful reheating and baryogenesis can take place in a large subset
of $D$-inflationary models.
\end{abstract}
\end{titlepage}
\setcounter{footnote}{0}
\setcounter{page}{1}
\setcounter{section}{0}
\setcounter{subsection}{0}
\setcounter{subsubsection}{0}


\section{Introduction} \label{sec:intro}

Successful slow-roll inflation requires that the inflaton potential be
sufficiently flat~\cite{slowroll}.  Since theories with
unbroken supersymmetry
naturally possess moduli spaces of exactly flat directions (the flatness
being preserved by quantum corrections), and because supersymmetric
theories are the theoretically most motivated extension of the Standard
Model (SM), supersymmetric theories of inflation have been much
studied.  In supersymmetric theories there are two possible
sources of a non-zero potential energy: $F$-term contributions
arising from terms in the superpotential, and $D$-terms
that arise from the supersymmetirization of the gauge kinetic energy;
either can in principle lead to the false vacuum density
necessary for inflation.

A significant problem for $F$-term inflation arises from the simple
fact that the non-zero energy density during the inflationary
period spontaneously breaks supersymmetry; moreover this breaking feeds
back into the inflaton potential, ruining the required
flatness~\cite{earlysusy,copeland,drt}.
Specifically, suppose the field whose non-zero $F$-term breaks
supersymmetry is $\psi$, with $\vev{\psi} = \theta^2 F_\psi$,
while the inflaton field itself is $\phi$.  No symmetry can prevent the
appearance of the term
\bea
\Delta {\cal L} &=& \frac{1}{\mpl^2}\int d^4\theta \phi^\dagger
	\phi \psi^\dagger \psi\nonumber \\
	&=& \frac{|F_\psi|^2}{\mpl^2} \phi^\dagger \phi ,
\label{eq:Fmass}
\eea
where, as suggested by supergravity, we have taken the reduced Planck
mass $\mpl = M_{\rm Planck}/\sqrt{8\pi}$ as the typical scale of
higher dimension operators. (We will also assume throughout that
supergravity is the dominant messenger of 
supersymmetry-breaking to the visible world.)
This effective mass squared term for $\phi$ is the same order of
magnitude as the Hubble constant squared, $H^2$, during inflation, since
$H$ is related to the total energy density, $\rho$, via
$H^2 = \rho/3\mpl^2$, and $\rho \simeq |F_\psi|^2$.  Thus it is
difficult to satisfy the slow-roll conditions for the inflaton field
\bea
\epsilon &\equiv & \frac{\mpl^2}{2}\left({d \ln V\over
d\phi}\right)^2 \ll 1,
\label{eq:slowroll1}
\\
|\eta| & \equiv & \frac{\mpl^2}{V} \left| {d^2 V\over
d\phi^2}\right| \ll 1,
\label{eq:slowroll2}
\eea
since $\eta$ picks up corrections of $\CO(1)$.
This problem is quite generic.  For instance in supergravity theories the
coupling in Eq.~(\ref{eq:Fmass}) naturally appears in the scalar potential
\beq
V = \exp\left( \frac{K}{\mpl^2} \right)
\left( W_i K^i_{\bar j} {\bar W}^{\bar j} - \frac{3|W|^2}{\mpl^2} \right)~+~
D\mbox{-terms},
\label{eq:sugraW}
\eeq
as a cross term between the K\"ahler potential
$K= \phi^\dagger \phi+...$ for the
inflaton field, and the expectation value of the
generalized $F$-term for $\psi$,
$W_\psi = \partial_\psi W + W \partial_\psi K /\mpl^2$.

However if it is assumed that $D$-terms provide the dominant contribution
to the vacuum energy~\cite{early}, then this problem is greatly ameliorated,
the basic reason being that a direct K\"ahler potential coupling of
the supersymmetry breaking $D$-term to the
inflaton direction, analogous to that for the $F$-term in
Eq.~(\ref{eq:Fmass}), is forbidden by gauge invariance.

In this paper we will be concerned with some issues that
are raised in constructing more fully realized models of $D$-term
inflation.  In Section~\ref{sec:model} we outline the basic
structure of $D$-term inflationary models, and we point out that the
constraints that such a model must obey are more severe than usually
assumed.  In Section~\ref{sec:reheat} we discuss reheating in $D$-term models
and how one should expect it to proceed. 

One of the most attractive aspects of $F$-term models of inflation is
the possibility of using the Affleck-Dine (AD) mechanism~\cite{ad} for
generating the observed baryon asymmetry.   In Section~\ref{sec:adb} we
argue that a variation of the AD baryogenesis schemes due to  
Murayama and Yanagida~\cite{hitoshi} and Dine, Randall and 
Thomas~\cite{drt} can be implemented in $D$-term inflationary models.
Section~\ref{sec:conc} contains our conclusions, and the Appendix discusses
in some detail the possible decoherence of the AD flat direction in a hot
environment.

\section{The basic model and constraints}\label{sec:model}

Consider a toy model of $D$-term inflation
\cite{halyobd,early}, whose field content includes, a neutral
chiral superfield, $S$, and superfields $\psi_\pm$ with charges
$\pm 1$ under a U(1) symmetry, U(1)$_{FI}$.  The superpotential
$W = \la S\psi_+\psi_-$,
leads to the tree-level scalar potential
\beq
V = |\la|^2  \left( |\psi_+\psi_-|^2 + |S\psi_+|^2 + |S\psi_-|^2 \right)
 +  \frac{g^2}{2}\left(|\psi_+|^2 - |\psi_-|^2 + \xi^2 \right)^2 .
\label{eq:toy}
\eeq
Here, a Fayet-Iliopoulos (FI) term, $\xi^2$, which we define to be
positive, has also been included.  The
supersymmetry preserving global minimum of this potential is at
$S=0$, $\psi_+ =0$, $\psi_- = \xi$.
However if $S$ is initially displaced sufficiently far from
its minimum, $S>S_{\rm crit}\equiv g\xi/\la$, then the local minimum
$\psi_+ =\psi_- = 0$ has non-zero energy density
$\rho = g^2\xi^4 /2$ that, to this order, is independent
of the value of the field $S$.  Most importantly, the flatness of
the potential for $S$ is not disastrously disturbed by the
inflationary epoch supersymmetry breaking, since the U(1)$_{FI}$ vector
superfield couples only to the K\"ahler potential of charged chiral
fields.  Moreover, it has been argued~\cite{lyth} that possible
dangerous non-minimal terms in the gauge kinetic function depending
on $S$ can quite easily be forbidden by a combination of symmetries
and holomorphy, although we will see in the following that the
problem is more serious than previously thought.

The leading inflation-induced curvature of the effective potential for
$S$ is usually assumed to be due to the superpotential couplings
between $S$ and charged chiral fields that couple in turn to the U(1)$_{FI}$
$D$-term.  In particular, at 1-loop, supersymmetry breaking leads
to the splitting in mass of
the bosonic components of $\psi_\pm$ from their fermionic
partners, and the potential for $S$ receives a correction
$\Delta V = C\alpha^2 \xi^4 \ln(|\la S|/\mu)$, where
$\mu$ is the momentum scale, $\alpha = g^2/4\pi$, and $C\ge 1$
counts the multiplicity of $\psi_\pm$ pairs in more general
toy models.
As a result the induced inflationary-epoch curvature of the
inflaton potential is suppressed relative to the Hubble constant
by small gauge coupling dependent loop factors.
Thus this model is an implementation of the hybrid
inflationary models of Refs.~\cite{hybrid}, where $S$ is the
field that slow rolls due to a nearly flat potential.

Actually, the situation is not quite so simple as just stated.
For the logarithmically dependent $S$ potential, $|\eta| \ll 1$
is the first of the slow-roll conditions to fail.  Solving for
the final value $S_f$ by imposing $|\eta(S_f)| = 1$ gives
$S_f =\sqrt{C\al \mpl^2/2\pi}$; for typical values of the parameters
this is one or two orders of magnitude larger than $S_{\rm crit}$.
Moreover, the slow-roll of $S$ must initiate at substantially larger
values if there is to be enough e-folds of expansion to solve
the flatness and horizon problems of the standard cosmology.
Requiring $N=55$ e-folds leads, under the assumption that the
logarithmic $S$ potential dominates, to an initial value
\beq
S_{55} = \left( {55 C\al \mpl^2 \over \pi} + S_f^2\right)^{1/2}.
\label{eq:initialS}
\eeq
For typical values of the gauge coupling, $S_{55}^2 \simeq 0.8 C \mpl^2$.
It is now necessary to ask if, given such large initial values, it
is reasonable to suppose that the dominant curvature of $V(S)$ is due
to 1-loop effects for the entire evolution of the field.  (Recall
in this regard that approximately 55 e-folds before the end of inflation
the fluctuations on the largest scales we currently observe were just
leaving the horizon; the observational bound on the spectral index
$n$ of these fluctuations is $|n-1| \lsim 0.2$.  Since the spectral
index of adiabatic density fluctuations is determined by the slow-roll
parameters evaluated at the appropriate epoch,
$n-1 = 2\eta_{55} - 6\epsilon_{55}$, deviations from a flat potential
can be important. In this paper we will not consider the CMBR
spectrum in detail because, among other effects, these models have 
cosmic string solutions which can significantly alter the power 
spectrum~\cite{jeannerot}.)

Consider, for example, the addition of a superpotential term of the
form 
\beq
\De W = \ka S^m/\mpl^{(m-3)}.
\eeq  
The resulting shift in $\eta$ is
\beq
\De \eta(S) \simeq 
\frac{2m^2(2m-2)(2m-3)\ka^2S^4}{g^2\xi^4} \left(\frac{S}{\mpl}\right)^{2(m-4)}.
\label{eq:etashift}
\eeq
Given the value $S_{55}\simeq \mpl$, the term in parentheses
is $\CO(1)$ and the expression is dominated by the large enhancement
$S_{55}^4/\xi^4 \sim 2\times 10^{10}$ (we will see shortly how the scale
of $\xi$ is set to be about $6.6\times 10^{15}\gev$ by CMBR measurements).  
Only for values of $\ka<10^{-5}$
do we get $\De \eta <1$. The smallest natural values we might expect for
$\ka$, if it is not identically zero by some symmetry, 
are $\ka\sim1/m!$. Such values might
arise by matching this non-renormalizable superpotential
coupling onto some underlying renormalizable theory in which heavy states have
been integrated out. Then the bound $\ka<10^{-5}$ requires that we must forbid
all operators of the form $S^m$ for at least $m\le 9$ in order to keep
$\De\eta<1$.
Note that increasing the multiplicity
factor $C$ just makes the problem worse.  

Further, as noted in Ref.~\cite{lyth}, there are also
dangerous non-minimal gauge-kinetic terms for U(1)$_{FI}$ of the form
\beq
\frac{S^k W^\al W_\al}{k!\mpl^k} +{\rm h.c.}
\label{eq:nonminf}
\eeq
The contribution of these operators to $\eta_{55}$ is in turn
\beq
\De \eta_{55}\simeq\frac{1}{(k-2)!}\left(\frac{S_{55}}{\mpl}\right)^{(k-2)},
\label{eq:etashift2}
\eeq
also $\CO(1)$ unless all operators with $k\lsim 6$ are forbidden.

The obvious way to forbid these two sets of operators while
preserving the operator $S\psi_+\psi_-$ is by
a discrete symmetry of sufficiently high order or by an
R-symmetry. The exact form of this symmetry will have a
significant impact on the couplings of $S$ to light MSSM fields, and
therefore on the reheat temperature resulting from decay of the coherent
oscillations of the $S$ field.  We will therefore address this issue
more fully in Section.~\ref{sec:reheat}.  

Note, however, that one
potential problem with a global symmetry such as R-symmetry is the
worry that they may be violated by quantum gravitational
effects~\cite{planck}.  The alternative of a discrete gauge
symmetry~\cite{DGT} might therefore be preferable, since such
symmetries are protected from violation by quantum gravity effects.
Moreover, string theory very commonly possesses such symmetries, in
contrast to the case of (exact) global symmetries.  

Apart from the aesthetic and theoretical appeal of forbidding the
dangerous $S^m$ and $S^k W^\al W_\al/\mpl^k$ operators by a {\it discrete}
gauge symmetry, there is another intriguing possibility that such a
structure allows.  Unlike the case of a continuous symmetry,
at some order the dangerous operators will eventually arise and the
potential will not be flat for $S$ greater than some value $S_W$.
We must require that the plateau at $S<S_W$ is long enough to get at least 
55 e-folds of inflation, which places a lower bound on $S_W$ ($S_W> S_{55}$)
and thus on
the order of the operator which lifts the $S$ potential. If, as traditionally
assumed, the potential is flat well past $S=S_{55}$, then the universe inflates
for many more than 55 e-folds and thus the spectral index $n$ is very close 
to 1 on all scales, up to and including superhorizon-size scales. This
very long period of inflation
simultaneously drives $\Om$ so close to unity that we should currently observe
a flat universe. However,
if $S_W$ is just such that the universe inflates for only 55 e-folds 
(\ie, $\ep(S)$ and $|\eta(S)|\geq1$ for $S> S_{55}$), then
$n$ is constant on small scales but deviates on scales comparable to our
current horizon. In particular, the non-trivial structure of the potential
at $S\simeq S_W$ is imprinted on the spectral index via the increase
of the functions $\eta(S)$ and $\ep(S)$ at $S\simeq S_W$. Since this
spectrum was imprinted at the very start of inflation, it is observable today
on the largest scales. This then
would be an example of what we might call ``just-so inflation''
in that we have a period of inflation just long enough to
solve the horizon and flatness problems, but {\it not so long}
as to have the current value, $\Om_0$, of $\Om=\rho/\rho_{\rm crit.}$,
approximate 1 {\it exponentially} closely.  In other words
with two sources of the $S$ potential -- the logarithmic
dependence from loop corrections, and the power dependence
from higher-dimension operators -- it only requires a discrete
choice of the order of the symmetry group to ensure that we
currently measure $\Om_0< 1$, but not very far away from $\Om_0=1$.
This ``just-so'' mechanism, based on two different types of
contribution to the potential of the inflaton, with one
controlled by a discrete symmetry, appears to be one of nicest
explanations of how an open universe could result from (and
be consistent with) an inflationary early universe.
(For implementations of open inflationary universes, see,
\eg~\cite{open}.)

Notwithstanding these issues, crucial to the formulation of
this model was the FI term $\xi$.  The most attractive origin
for this term is the so-called anomalous
U(1)$_X$ symmetry of some string compactifications, whose apparent
field-theoretic anomalies are cancelled by a 4-dimensional version of the
Green-Schwarz mechanism involving shifts in the model-independent
axion~\cite{fiterm}.
In such a compactification a FI term is automatically generated, and its
magnitude has been calculated to be~\cite{fiterm}
\beq
\xi^2 = {\Tr(Q_X)\over 192 \pi^2} g_{str}^2 \mpl^2.
\label{eq:FI}
\eeq
Here the model-dependent factor $\Tr(Q_X)$ is proportional to the
total gravitational anomaly; in typical semi-realistic string models
$\Tr(Q_X)\sim\CO(50)$.
Now, the magnitude of the FI term sets the size of the Hubble constant
during inflation, and in turn the magnitude of the fluctuations in the
cosmic microwave background radiation (CMBR).  Normalizing to the
observed CMBR fluctuations requires~\cite{LR},
$\xi_{\rm CMBR} = 6.6\times 10^{15}\gev$, independent of $g$.
(This assumes that the logarithmically dependent $S$ potential
receives no significant corrections at $S_{55}$.  In the
general case $(V_{55}/\epsilon_{55})^{1/4} = 6.7\times 10^{16}\gev$.)
Unfortunately this is smaller than the prediction, Eq.~(\ref{eq:FI}),
of string theory by a sizeable amount.  Although some improvement
of this problem is possible by increasing the 1-loop coefficient $C$
in $\Delta V(S)$, it is argued in Ref.~\cite{LR} that the two values
cannot be brought into agreement unless the string prediction is
lowered, or some other mechanism for generating $\xi$ (without
simultaneously large $F$-terms) is found.  At present this is an open
problem.

One possibility in this regard is the suggestion that the
expectation value of the dilaton, which sets the strengths of
all couplings, is shifted during the inflationary epoch from its
current value, thus allowing a significantly smaller
$\xi$~\cite{matsuda}.  To agree with the CMBR normalization,
an inflationary value of $g_{\rm str.}\sim 1/50$ would be needed.
It is interesting to note in this
context that reducing the effective value of $g$ during inflation
also helps alleviate the problem with the higher-dimensional
superpotential and gauge-kinetic functions, as can be seen by
counting powers of $g$ in the expression Eq.~(\ref{eq:initialS})
for $S_{55}$.  In the following we will just {\it assume}
that some variant leads to the correct value of $\xi$ (which we take
to be $6.6\times 10^{15}\gev$), and focus on
the post-inflationary phenomenology of reheating and baryogenesis in $D$-term
models.

\section{Reheating}\label{sec:reheat}

Even if the magnitude of $\xi$ can be made consistent with the CMBR
data, successful reheating is still non-trivial.
First there are the usual constraints on the reheat
temperature arising from the overproduction of gravitinos~\cite{moroi}, 
although these constraints are somewhat modified in $D$-term models, since 
there are typically at least two well separated stages of reheating. In 
addition, the particular mechanism by which one gets sufficient baryogenesis
is greatly influenced by the details of reheating.

After inflation ends, the vacuum energy that drove inflation moves to
potential and kinetic energy for condensates of fields oscillating around
their various minima.  In the model outlined above, the original vacuum
energy, $V=\frac{1}{2}g^2\xi^4$, is now being carried by some combination of
the $\psi_-$ and $S$ fields.  The relative portion of the energy in each
oscillating condensate depends on their effective masses after inflation,
which are $\la\xi$ and $\sqrt{2}g\xi$ respectively.
For $\la^2=2g^2$ they share the energy evenly; we will assume
that $\la\sim g$ so that the contributions of the two condensates
to the total energy density of the universe are of the same order.
Moreover, because both condensates have quadratic
potentials, their energy densities will both scale as matter (\ie, $R^{-3}$)
as the universe expands, and so the ratio of their energy densities remains
constant.

The universe remains cold until $H$ drops below the decay width of
either one of the $S$ or $\psi_-$ fields, at which point that field
will decay, dumping its stored energy and reheating the universe.
At a later time the second condensate decays, dumping further entropy
and diluting any products of the first stage of reheating.
The amount of reheating is controlled by the widths
of these fields, which in turn depend on their masses and couplings.
In particular, at the reheat time, $t=t_R$, when $H\sim\Ga_A$ (here,
$A$ is a generic condensate field), the decay of the $A$ field dumps
energy density
\beq
\rho_A(t_R)=3\Ga_A^2\mpl^2 f_A
\eeq
into the vacuum, where $f_A$ is the fraction of $\rho_{\rm tot}$ stored
in the $A$ condensate.  The corresponding reheat temperature is
\beq
T_R\simeq\left(\frac{30\rho_A(t_R)}{\pi^2 g_*(t_R)}\right)^{1/4}
\simeq 0.4 f_A^{1/4}\sqrt{\Ga_A\mpl}.
\eeq
where $g_*(t_R)$ is the number of relativistic degrees of freedom at
$t=t_R$; one expects that $g_*(t_R)\sim {\rm few}\times 10^2$.

What are the bounds on $T_R$?  There are a number of model-dependent
limits; for example, in a GUT one requires $T_R\ll M_{\rm GUT}$ in
order to avoid creating stable monopoles.
However there are more stringent bounds coming from gravitino
production in supergravity theories.  For unstable gravitinos
in the mass range $100\gev$ to $1\tev$, one requires
$T_R \lsim 10^{7-9}\gev$~\cite{moroi}.  Also there is lower
bound on $T_R$ of approximately $6~\mev$ due to the fact that
the universe must reheat sufficiently that conventional Big Bang
Nucleosynthesis (BBN) is possible~\cite{bbn}.  Note that the quoted
gravitino bound on $T_R$ assumes that there is no later entropy dump
which dilutes the gravitino number density.  In $D$-term models
there are a number of stages of reheating, each with its own
entropy release, so this dilution must be taken into account.

The amount by which the gravitino relic of the first stage of reheating
is diluted by the second stage is easily calculated.  Suppose a (generic)
$X_1$ condensate decays at $H_1\simeq \Ga_1$
(at which time, for simplicity, we take $\rho_1 \simeq \rho_2$)
with reheat temperature $T_1(t_1)$, while the $X_2$ condensate decays
much later at time $t_2$ when $H_2\simeq \Ga_2$.
Since the energy density in the $X_2$ condensate
scales like non-relativistic matter $\rho_X \sim R^{-3}$, it is easy to
show that at the epoch of $X_2$ decay the radiation temperature has fallen
from $T_1(t_1)$ to
$T_1(t_2) = T_1(t_1)(\Ga_2/\Ga_1)^{2/3}$.  Thus the ratio of
entropy densities just before $X_2$ decay (when the
entropy is almost entirely in the form of red-shifted radiation
from $X_1$ decay) to that just after is given by
\bea
{s_{{\rm before},2} \over s_{{\rm after},2}}
&=& { g_{1*} (\Ga_1^2 \mpl^2/g_{1*})^{3/4} (\Ga_2/\Ga_1)^2 \over
g_{2*} (\Ga_2^2 \mpl^2/g_{2*})^{3/4} }\nonumber
\\
&\simeq& \sqrt{ \Ga_2 \over \Ga_1 } \simeq
{  T_2(t_2) \over T_1(t_1)},
\label{eq:sscaling}
\eea
where $g_{1*} \equiv g_*(t_1)$, \etc.
(In the last two equalities we have dropped the weak dependence
on the change of the effective number of relativistic degrees
of freedom, $g_*$.)  In the case of a {\it single} stage of
reheating, the yield of gravitinos, $Y_{3/2} \equiv n_{3/2} / s$,
is well approximated by the simple linear form,
$Y_{3/2}(T \ll 1~\mev) \simeq 2.14\times
10^{-11} (T_R/10^{10}\gev)$~\cite{moroi}.
This, together with Eq.~(\ref{eq:sscaling}), shows that the
diluted gravitino yield from the first epoch of reheating is in our case
{\it equal} to the yield from the second epoch of reheating (up to
$\CO(1)$ coefficients); \ie, just after $t_2$,
\beq
{s_{{\rm before},2}\over s_{{\rm after},2} }\,
Y^{(1)}_{3/2} \simeq Y^{(2)}_{3/2}.
\label{eq:gravyield}
\eeq
Thus in the models we are discussing,
the gravitino constraint on the reheat temperature should be applied to
the later of the two epochs of coherent field oscillation decay.

In any case, since the post-inflationary masses of both $S$ and
$\psi_-$ are large (of order of $\xi\simeq 6\times10^{15}\gev$),
even small couplings can lead to untenable reheat temperatures.
Thus it is necessary to investigate the couplings of $\psi_-$
and $S$ to light fields, given the constraints of $D$-term
inflation.

\subsection{Immediate decay of $\psi_-$ condensate}

The most constrained decays involve the $\psi_-$ oscillations.
Let us first consider the case in which the non-zero FI term is generated
through cancellation of the anomaly of a pseudo-anomalous U(1)$_X$.
Recall that in order for the four-dimensional version of the Green-Schwarz
mechanism to operate, there must exist
non-zero mixed U(1)$_X$G$_A^2$ anomalies with all gauge groups G$_A$.
In particular there must exist fields that are simultaneously charged
under both U(1)$_X$ and each of the subgroups of the Standard Model,
and that are chiral with respect to this combination of groups.
Therefore, the anomalous $D$-term {\it must}\/  receive contributions
from fields which are charged under G$_{\rm SM}$,
and which have $\CO(1)$ charges under U(1)$_X$.
The $D$-term therefore has the form:
\beq
D=g\left(|\psi_+|^2-|\psi_-|^2+\sum_i q_i|Q_i|^2+\xi^2\right)
\eeq
where the $Q_i$ are fields charged under $G_{SM}$ which have charge $q_i$
under the anomalous U(1)$_X$.  (The charges must satisfy
$q_i>0$ so that at the post-inflationary minimum none of the
fields $Q_i$ gain expectation values of order $\xi$, since otherwise the
Standard Model gauge group is broken at an unacceptably high scale.)
If the usual MSSM fields, $\phi_i$, 
are included among the $Q_i$ then they couple to
fluctuations of the $\psi_-$ field around its expectation
value with strength
\beq
\Delta \CL = g^2 \xi \sum_{FI} q_i |\phi_i|^2 (\delta \psi_-
+ \delta\psi_-^\dagger).
\label{eq:cubic}
\eeq
The resulting decay
width for $\psi_-$ is $\Ga_{\psi_-}\sim g^3\xi/16\pi$ and so
$\psi_-$ immediately decays, giving $T_R\sim
10^{15}-10^{16}\gev$. From our earlier discussion of two-stage reheating, 
it is clear that this violates the gravitino bound {\it unless}\/ the 
$S$ field decays with $T_R<10^9\gev$. 

Moreover, such a large $\psi_-$ reheat temperature does lead to the evaporation
of the $S$-condensate into an incoherent collection of $S$-particles. Recall 
that the MSSM fields $\phi_i$ and the $\psi_+$ interact via exchange of
the U(1)$_X$ gauge field. After U(1)$_X$ breaks via $\vev{\psi_-}=\xi$, the
gauge field can be integrated out, leaving an effective K\"ahler potential
term of the form 
\beq
\Delta K=\frac{\phi^\dagger\phi\,\psi_+^\dagger\psi_+}{\xi^2}.
\label{eq:delK}
\eeq
But because of the form of the superpotential ($W=\lambda S\psi_+\psi_-$),
$\vev{F_{\psi_+}}=\lambda S\vev{\psi_-}=\lambda \xi S$. Eq.~(\ref{eq:delK})
then produces an interaction in the effective Lagrangian between $\phi_i$
and $S$ scalars:
\bea
\Delta {\cal L}&=&\frac{|\phi_i|^2 |F_{\psi_+}|^2}{\xi^2} \nonumber \\
&=&\lambda^2|\phi|^2 |S|^2.
\label{eq:Leff}
\eea
The corresponding scattering rate for an $S$-scalar in a bath of MSSM fields
at temperature $T$ is then
\beq
\Gamma_{\rm scatt}=n_\phi \vev{v\sigma}\sim\frac{\lambda^4}{16\pi T^2}T^3.
\eeq
If $\psi_-$ decays immediately after the end of inflation, as is the case
if the MSSM fields have ${\cal O}(1)$ U(1)$_X$ charges, 
then $H\sim\sqrt\rho/\mpl\sim T^2/\mpl\ll T$ since $T\lsim\xi$.
Thus $\Gamma_{\rm scatt}\gg H$ and the $S$-condensate is evaporated into an
incoherent collection of $S$-particles which then later decay at $t\simeq
\Gamma_S^{-1}$, where $\Gamma_S$ is the $S$ decay 
width.~\footnote{Implicit in this calculation is that the MSSM fields are
relativistic; that is, they do not receive masses $\sim T$. Such masses
do not arise from the $D$-term simply because it is cancelled off by 
$\vev{\psi_-}$. Similarly, such large masses for the first two MSSM generations
cannot be generated by some large AD condensate vev because of their small
Yukawa couplings.} As long as the individual $S$-particles decay with
reheat temperature $\lsim 10^9\gev$, the fact that the condensate has been
destroyed does not violate the gravitino bounds. We will return to 
$S$-decays in Section~\ref{sec:Sdecay}. We will also see in 
Section~\ref{sec:adb}\ that this very high reheat temperature 
for $\psi_-$ can also disrupt certain implementations of AD baryogenesis.

\subsection{Delayed decay of $\psi_-$ condensate through kinetic mixing}
\label{sec:km}

There is one way to satisfy the anomaly conditions for anomalous U(1)$_X$
models which does not lead to immediate $\psi_-$ decay:
assume that the $Q_i$ are all vector-like with respect to
$G_{SM}$, but chiral with respect to U(1)$_X$. Thus the fields may naturally
acquire masses of $\CO(\xi)$, and $\psi_-$ oscillations can no longer decay
into them.  (Treating the decay of the condensate as the decay of a
collection of individual condensate particles.   Solutions to the coupled
$S$, $\psi_-$ equations of motion
corresponding to parametric resonance do in principle allow for decay
of a condensate into heavier fields, but recent work~\cite{param}\
indicates that this is in fact very difficult to achieve in an
expanding universe.) It may also be that the FI term which drives 
inflation arises from a non-anomalous U(1)$_{FI}$, in which case there is
no need for {\it any}\/ fields charged under the U(1)$_{FI}$ to also carry 
MSSM charges.

This would
appear to be an acceptable solution: the U(1)$_{FI}$ can drive inflation
but decouples from all low-energy physics since no light field can be
charged under it.  This is not what one might have hoped (since there are,
for example, 
many interesting models which use pseudo-anomalous U(1)'s to explain low-energy
phenomena), but it at least seems consistent.

In fact it may not be, since we have ignored an important effect that
seems unavoidable in models with FI terms. This effect is
kinetic mixing between the U(1)$_{FI}$ and hypercharge, U(1)$_Y$, gauge
symmetries~\cite{holdom,dkmr}.  Recall that
in a theory with two U(1) factors, the most general renormalizable
Lagrangian contains a gauge-invariant term which mixes the gauge field
strengths of the two U(1)'s.  In the
basis in which the interaction terms have the canonical form, the
pure gauge part of the supersymmetric Lagrangian for an arbitrary
U(1)$_a\times$U(1)$_b$ theory is
\beq
\CL_{\rm gauge}=\frac{1}{32}\int d^2\th\,\left\{W_aW_a+W_bW_b-2\chi W_aW_b
\right\}
\eeq
where $W_a$ and $W_b$ are the chiral gauge field strength superfields for
the two gauge symmetries ($W=\bar D^2DV$ for a vector superfield $V$).
(We assume in the following that the kinetic mixing parameter $\chi\ll 1$.)
To bring the pure gauge portion of the Lagrangian to canonical form,
a shift of one of the gauge fields is necessary:
\beq
V^\mu_b\to V'^\mu_b=V^\mu_b - \chi V^\mu_a
\eeq
which implies $W_b\to W'_b=W_b-\chi W_a$.
This particular basis is dictated by the assumption
that U(1)$_a$ is broken by the expectation value of some field with
non-zero U(1)$_a$ charge.
Thus in the case of interest, we identify U(1)$_a$
as the U(1) with the FI term, U(1)$_{FI}$, and U(1)$_b$ as U(1)$_Y$.

In this basis the gauge Lagrangian is diagonal,
however, the interaction piece is modified.   In particular
suppose initially that there was a sector of fields
$\Phi_i$ charged only under U(1)$_{FI}$, and another sector
of fields $\phi_i$ charged only under U(1)$_Y$ (the MSSM fields
in our case), then the basis change introduces
a number of new interactions between the two sectors.  Most
importantly for the present purposes, upon solving for
the $D$-terms, one finds~\cite{dkmr}:
\bea
D_{FI}&=&-g\sum_i q_i |\Phi_i|^2-g_y\chi\sum_i y_i
|\phi_i|^2 + \xi^2 \nonumber \\
D'_y&=&-g_y\sum_i y_i |\phi_i|^2,
\label{eq:dterms}
\eea
namely, that the U(1)$_{FI}$ $D$-term picks up an admixture of the
hypercharge $D$-term (here $y_i$ is the hypercharge of the various
fields).

As discussed in detail in Refs.~\cite{holdom,dkmr},
non-zero $\chi$ can
arise from threshold effects, either field-theoretic or string
theoretic, or renormalization group running of the couplings in
the low-energy field theory.  The general requirement that this
occurs is that there exist particles charged under both U(1)'s
such that, in an effective field theory language where we
integrate out modes once we drop below their
mass, $\Tr(Q_a Q_b) \neq 0$ for some range of momentum scale.

There are two symmetries that naturally forbid the existence of kinetic
mixing.  The first is simply gauge invariance if one of the U(1)'s
actually sits inside an unbroken non-Abelian group (and in this case
$\Tr(Q_a Q_b) = \Tr(Q_a)\Tr(Q_b) = 0$ for each, degenerate, mass level).
The second symmetry that forbids kinetic mixing is an unbroken
``charge conjugation'' symmetry that acts on {\it only one} of the
two U(1)'s, $A_\mu \to -A_\mu$, which again enforces
the vanishing of the trace~\cite{dkmr}.

However in the case of U(1)$_Y$ mixing with U(1)$_{FI}$,
neither of the above symmetries can be exact.
In particular, charge conjugation is certainly not an unbroken
symmetry of the hypercharge interaction, and, most importantly,
 a charge conjugation symmetry acting on 
U(1)$_{FI}$ {\it would also forbid the FI term itself}.  Moreover, in the
case of the pseudo-anomalous U(1)$_X$ symmetry, there must exist
particles charged under both U(1)'s.
Therefore non-zero kinetic mixing is typically generated in
the case of interest.

Given that $\chi$ is not exactly zero due to some unbroken symmetry,
the natural expectation for $\chi$ is that is it generated by a one-loop
threshold effect and thus $10^{-3} \lsim \chi \lsim 10^{-2}$.  This
estimate is applicable to string theoretic threshold corrections as well
as field theory threshold effects~\cite{dkmr}.  A reduction in
the value of $\chi$ generated by loop effects is possible if all
states simultaneously charged under U(1)$_Y$ and U(1)$_{FI}$ are, as far
as their quantum numbers are concerned, effectively in multiplets of
some non-Abelian group that contains one of the U(1)'s.  The fact that
this non-Abelian gauge symmetry must be broken implies that the
states in the multiplets are not naturally degenerate, and thus $\chi$
will again be generated, but with a value now suppressed by the
splitting in the masses of the states relative to their common mass.
In this way $\chi$ can be reduced to a 2 or 3-loop effect (if,
respectively, one or both sets of U(1) quantum numbers embed
in some effective non-Abelian multiplet): $10^{-6}\lsim\chi\lsim10^{-4}$.

The end result, therefore, in the case of $D$-term inflation,
is that the full potential contains
a contribution from the modified U(1)$_{FI}$
$D$-term, now including an admixture of the usual hypercharge
$D$-term, as well as the hypercharge $D$-term itself:
\beq
V_D = \frac{g^2}{2} \left(  |\psi_+|^2   - |\psi_-|^2 +
\chi\sum_i y_i |\phi_i|^2 +
\xi^2\right)^2 + \frac{g_y^2}{2}
\left(\sum_i y_i |\phi_i|^2\right)^2.
\label{eq:potentialnew}
\eeq
Here, $\phi_i$ are the usual MSSM fields of hypercharge $y_i$ (we have
dropped contributions from any heavy fields which are vector-like under
with respect to the MSSM since they play no role in the following).
The presence of the extra terms in the U(1)$_{FI}$ $D$-term
does not destroy the possibility of inflation in this model.
For $\vev{S}$ sufficiently large, the local minimum is still at
$\vev{\psi_+}=\vev{\psi_-}=0$, as a consequence of the presence of the
hypercharge $D$-term, while the expectation values of the MSSM
fields $\phi_i$ satisfy the constraint
\beq
\sum_i y_i \vev{|\phi_i|^2} = \frac{ \chi g^2 \xi^2/4}{{\chi^2}
g^2/4+g_y^2 }.
\label{eq:Lexpectation}
\eeq
At this local minimum the combined $D$-terms are
non-zero, and the energy density is
\beq
\< V \> = \frac{g^2 g_y^2 \xi^4 }{{{\chi }^2}
g^{2}/2+2 g_{y}^{2}}.
\label{eq:VwithKM}
\eeq
Moreover, at the true post-inflationary minimum we still have
$\vev{\psi_-}=\xi$, despite the fact that the U(1)$_{FI}$ $D$-term now
inevitably contains fields other than $\psi_-$ with negative
charge. Thus the alteration of the potential imposed by kinetic mixing
still permits inflation as before.

However, kinetic mixing 
now allows $\psi_-$ to decay to pairs of light MSSM scalars 
through a coupling of strength $g^2\chi\xi$. Note that such mixing
allows $\psi_-$ to couple directly to light MSSM fields even if no such 
coupling existed in the bare Lagrangian.

In Section~\ref{sec:adb}
we will see that AD baryogenesis occurs at $T\simeq 10^{10-11}\gev$. We have
already argued that the natural range for $\chi$ is $10^{-6}\lsim\chi
\lsim 10^{-2}$. Over most of this range, $\psi_-$ decays before AD 
baryogenesis, \ie, $T_R^{(\psi)}>T_{AD}$, and the resulting physics is very
similar to the immediate decay scenario outlined in the previous section.

However if
$\chi\lsim 10^{-6}$, then $\psi_-$ decays after baryogenesis, allowing
further implementations of the AD scenario.
How would $\chi\lsim 10^{-6}$ arise in a realistic model?
First, if hypercharge
is embedded in a GUT at the unification scale (rather than string
unifying at some possibly higher scale), there are no
significant renormalization group enhanced contributions to
$\chi$; only threshold corrections to the initial (zero) value of
$\chi$ are relevant.  Because of the tracelessness of the hypercharge
for any GUT multiplet, the GUT symmetry can naturally
reduce these contributions to $10^{-4}$, the only exceptions
being if split multiplets, such as the GUT Higgs
$({\bf 5} + {\bf \bar5})$, which contain light doublets
and heavy triplets, are charged under U(1)$_{FI}$.  However from
our earlier analysis we know that no light MSSM states, such as the
doublets, can have ``bare'' U(1)$_{FI}$ charges, and so this exception
does not arise.  To further reduce $\chi$ we can impose an
(approximate) ``charge-conjugation'' symmetry.
For instance, if the spectrum of states charged under U(1)$_{FI}$
is of the form $({\bf 5}_q + {\bf \bar 5}_q)$, where the subscript
is the charge, then so long as the states are approximately
degenerate in mass the threshold induced $\chi$ is further
suppressed. In this way, a value of $\chi\lsim 10^{-6}$ is possible.
However, it seems very difficult to imagine values for $\chi$ very 
far below this limit.

\subsection{Decay of $S$} \label{sec:Sdecay}

We now turn to the fate of the $S$ field,
which depends upon its available decay modes. The $S$ and
$\psi_+$ fields are nearly degenerate, while the mass of the $\psi_-$
field is suppressed/enhanced by $g/\la$.  In either case, if $S$ is
kinematically allowed to decay to $\psi_+$ or $\psi_-$,
it will do so with $\Ga_S\sim\xi$, producing $T_R\sim 10^{15-16}\gev$.
We already argued that $\psi_-$ is likely to decay early; since we cannot
have both $\psi_-$ {\it and}\/ $S$ decaying quickly, we conclude that
$\Ga_S\sim\xi$ violates the gravitino bound. (The later decay of the
AD condensate cannot dump enough entropy to avoid this conclusion.)
Ruling out these decays kinematically, one is left with decays of $S$ to 
light matter. However, the constraints
discussed in Section~\ref{sec:model} on
superpotential terms of the form $S^k$, or non-minimal
gauge-kinetic terms of the form $S^m W^{\al} W_\al$,
indicate that the couplings of $S$ are strongly restricted by
symmetries.  As mentioned in Section~\ref{sec:model} an attractive
choice is to use R-symmetry to limit the couplings of $S$.

Thus consider $R(S)\neq 0$.  Since the usual
MSSM superpotential has R-charge $R(W_{\rm MSSM}) = 2$, no
superpotential couplings are allowed between
$W_{\rm MSSM}$ and $S^k$ (at least until $R$ is broken at an effective
scale $\CO(\mgrav)$ within the visible sector).
However, $S$ can couple to $r$th-order invariants
composed of MSSM fields for $r>3$.
For example, if we assign continuous $R$-charges
$R(S)=-2$, $R(\psi_\pm)=2$, and all non-Higgs MSSM fields $R = 1$, then
the there exists the term $SQQQL/\mpl^2$ in the superpotential.
Alternatively, if we take $R(Q,L,u^c,d^c,e^c) = 1/2$ and $R(H_u,H_d)=1$,
then the term $S(H_uH_d)^2/\mpl^2$ is the lowest-dimension operator.
For general $r$ the leading superpotential terms
involving $S$ are schematically of the form
\beq
W = \la S\psi_+\psi_- + \ka \frac{S\phi^r}{\mpl^{(r-2)}},
\label{eq:gensc}
\eeq
where the $\ka$ coupling leads to a $r$-body decay of the scalar
component of $S$ to $(r-2)$ scalars and two fermions.
The resulting visible sector reheat temperature is
\beq
T^{(S)}_R \simeq { \ka \la^{(r-3/2)} \over (3\times 10^{2})^{(4-r)} }
\left(\frac{P_4}{P_r}\right)^{1/2}\times(4\times 10^{8}\gev),
\label{eq:sreheat}
\eeq
where $P_4/P_r$ is the ratio of the 4-body phase space factor to
the $r$-body, and we have taken $\xi=6.6\times10^{15}\gev$. 
If $\ka,\la$ are $\CO(1)$, then for $r=4$ the reheat temperature is
at the upper end of the gravitino bound; for $r>4$, $T_R^{(S)}$ is well within
the constraints imposed by gravitino production. Notice that for all
$r\geq4$ the decay of the $S$-fields occurs {\it after}\/ the epoch of
AD baryogenesis at $H\simeq\mgrav$. (Because we assume supergravity mediation,
$\mgrav\simeq\mw\simeq1\tev$.)

Also note that such an R-symmetry acting on $S$ and
$\psi_\pm$ has the further advantage of explaining both the absence
of the $S^m W^{\al} W_\al$ operators, and
a direct mass term $M\psi_+\psi_-$.  Such a mass term, if taken
to be $\CO(\mpl)$, would lead to the breaking of supersymmetry at an
unacceptably high scale at the end of inflation. We have also checked that, 
for $r=4$, K\"ahler potential couplings of $S$ to MSSM fields
never parametrically dominate over the superpotential
interactions, and thus the reheat temperature is naturally maintained in
the $10^8\gev$ range.

~

In summary, for generic U(1)$_{FI}$ charges and couplings, 
we expect the $\psi_-$ 
condensate to decay immediately after the end of inflation, reheating the
universe to a temperature $T_R^{(\psi)}\simeq 10^{15}\gev$. The bath of light
MSSM fields scatter off the $S$ condensate, evaporating it into a plasma of
incoherent $S$ scalars. These scalars, however, must not decay until
$H\simeq\Ga_S\lsim1\gev$ to give a final $T_R\lsim10^9\gev$ 
consistent with the gravitino production bounds. We have shown that given
the $R$-symmetries necessary for successful inflation, such small widths
for $S$ are actually quite natural.
As we will see in the next section, baryogenesis occurs at $H\simeq1\tev$ 
and thus $S$-scalars decay after
baryogenesis is complete. There is also the possibility that $\psi_-$ 
has no bare couplings to light MSSM fields in which case its decays will
be dominated by kinetic mixing effects. Here, if $\chi\gsim10^{-6}$, the
thermal history is identical to that described previously. But if $\chi\lsim
10^{-6}$ then neither of the coherent $S$ nor $\psi_-$ condensates 
decays or evaporates until after baryogenesis has
already occurred.

\section{Affleck-Dine baryogenesis}\label{sec:adb}

We now turn to the question of whether AD baryogenesis
can be successfully implemented in the $D$-term inflationary
scenarios.  We will argue that an adaption of the versions of AD 
baryogenesis studied by Murayama and Yanagida~\cite{hitoshi}\ and
Dine, Randall and Thomas~\cite{drt}\ can lead
to successful baryogenesis in the case of $D$-term inflation.

In general we will not
make a distinction between direct AD baryogenesis and baryogenesis
{\it via} leptogenesis.  As long as the temperature
of the SM degrees of freedom after the decay of the AD condensate
is greater than the weak scale, non-perturbative sphaleron processes
efficiently convert lepton number to baryon number, destroying
$(B+L)$, but leaving asymmetries in both $L$ and $B$ as long as
non-zero $(B-L)$ was also produced by the evolution of the condensate
(as is typically the case).

Let us recall the basic idea of the implementations of
AD baryogenesis as outlined in Refs.~\cite{hitoshi,drt}.  
First, an MSSM flat direction lifted by a
(baryon or lepton number violating) higher-dimension
operator takes on a large expectation value in the {\it post-inflationary}
universe due to a negative effective mass-squared term, proportional to
$H^2$ (coming from terms similar to that discussed in Eq.~(\ref{eq:Fmass})).
Due to inflation, the phase of this condensate is uniform over a region
at least as large as our currently observable universe.
When this $H$-dependent mass drops below the usual
$T=0$ supersymmetry breaking soft mass of the flat direction (\ie,
when $H\lsim \mgrav$), the condensate begins to oscillate.  Furthermore,
at this epoch, the magnitude of the baryon or lepton number violating
$A$-terms in the potential (proportional to the $B$ or $L$ violating
terms in the superpotential) are of the same magnitude as the
$B$ or $L$ conserving terms. The relative phase between the $A$-terms
and the condensate is a source of CP-violation
that can be naturally large.  In such a situation 
the oscillating condensate picks up a $B$ or $L$ number per
mode of $\CO(1)$.  Finally the AD condensate decays in a $B$
or $L$ conserving way, releasing these quantum numbers into
light MSSM degrees of freedom.
Note, that for this mechanism to apply, it is {\it not}
necessary for the AD condensate to have a large expectation
value either during, or immediately after, inflation.  It
is only important that at $H\simeq\mgrav$ it have a large vev and 
well-defined phase over volumes comparable to the currently observable 
universe.

As enumerated in Refs.~\cite{drt,dflat}, there are many
possible flat directions
in the MSSM that could be involved in AD baryogenesis.  We will
label a generic MSSM flat direction by $\ph$.  Examples of
such flat directions are provided by $LH_u\sim\ph^2$ (in
terms of doublet components: $L=(\ph,0)$, $H_u=(0,\ph)$), or by
$Q_1 L_1 {\bar d_2}\sim\ph^3$.

The magnitude of the 
expectation values of the MSSM flat directions after inflation
are primarily fixed by higher dimension operators in the K\"ahler
potential that couple the flat direction to other fields.  In
particular consider the couplings
\beq
\De \CL = \int d^4\th\,  (c_1 S^\dagger S + c_2 \psi_+^\dagger \psi_+
+ c_3 \psi_-^\dagger \psi_- +\cdots) {\ph^\dagger \ph\over \mpl^2},
\label{eq:kcouplings}
\eeq
where the $c_i$'s might be expected to be $\CO(1)$ constants of either
sign.  There is no symmetry of the low-energy effective
theory that can forbid these couplings from appearing.
In terms of components, Eq.~(\ref{eq:kcouplings}) includes the terms,
$[ c_1 (|\p S|^2 +|F_S|^2) + c_2 (|\p \psi|^2 + |F_\psi|^2)
+ ...] {|\ph|^2 / \mpl^2}$, involving both the kinetic and potential
energies of the various degrees of freedom.  The only contribution to the
energy that is not included in this direct coupling is that arising
from $D$-terms, since such a coupling is disallowed by gauge invariance.
In the case where all chiral superfields in the model couple with
equal strength, $c_i = c$, Eq.~(\ref{eq:kcouplings}) couples the total
energy density, minus the $D$-term contribution, to $|\ph|^2$, leading
to an effective mass-squared $-c(\rho_{\rm tot}-\rho_{\rm D})/\mpl^2$
for the scalar component
of $\ph$. Of course, in general, there is no symmetry
that guarantees that all the $c_i$'s are of the same magnitude,
or even sign, but in the case of interest to us, only the 
$S$ field (and perhaps $\psi_-$ if it is long-lived) contain any
significant energy density in the post-inflationary universe.
As long as $c_S$ (and perhaps $c_{\psi_-}$ also~\footnote{Unequal 
coupling coefficients $c_i$ cause oscillations around some mean value, and
can potentially lead to quite complicated behavior at intermediate times,
where energy is exchanged between the $S,\psi_-$ and AD modes.}) 
is of the appropriate sign
and $\CO(1)$, then it is a reasonable approximation to take
$|\ph|^2$ to couple to $(\rho_{\rm tot}-\rho_{\rm D})$.

If we take the leading non-renormalizable (and $B-L$ violating)
superpotential term that lifts the flat direction to be
$h\ph^n/\mpl^{n-3}$, the structure of the scalar potential
for the flat directions is then
\beq
V = \left(\mgrav^2 + \De m^2 \right) |\ph|^2
+ \frac{\mgrav}{\mpl^{n-3}}(A h \ph^n +
{\rm h.c.}) + {|h|^2 |\ph|^{2(n-1)} \over \mpl^{2(n-3)} } 
\label{eq:flatV}
\eeq
This includes the explicit soft mass-squareds coming from supersymmetry
breaking,
the $|\ph|^{2(n-1)}$ term arising from the leading non-renormalizable
superpotential term, the associated $A$-terms for supergravity
mediation (with the appropriate power
of the gravitino mass $\mgrav$ factored out), and, importantly, the
$\De m^2 |\ph|^2$ term from expanding out the non-minimal K\"ahler
potential Eq.~(\ref{eq:kcouplings}).  Notice that there are no 
$A$-terms proportional to $H$ arising from the finite energy density of
the universe after inflation. In an $F$-inflationary scenario, such terms
do typically arise, but in $D$-inflation the form of the superpotential
does not produce terms of this type because $\vev{\psi_+}=0$ and thus
$\vev{W}=0$ at all times.

Lacking any other information, it is natural to expect $h$ in 
Eq.~(\ref{eq:flatV}) to be $\CO(1)$.
In particular, there is one non-renormalizable term about which we
possess some experimental information.
As noted in Refs.~\cite{hitoshi,drt}, the operator that lifts the
$LH_u$ flat direction, $h(LH_u)^2/\mpl$, is the same MSSM operator that
gives rise to Majorana masses for the left handed neutrinos.  Thus,
in this case, the coefficients $h$ are related to the
neutrino mass spectrum.  To be precise, there are three $L_iH_u$ flat
directions ($i=e,\mu,\tau$), and the superpotential operator is
$h_{ij} L_i^a L_j^b H_u^c H_u^d \epsilon_{ac} \epsilon_{bd}$, with
$a,b,c,d$ SU(2) indices.  Since the flatter the AD potential
the larger the resulting baryon asymmetry, it is the smallest
of the couplings in the potential that dominate.  If $m_{ij}$ is the
neutrino Majorana mass matrix then the term in the potential
for the $L_i H_u \equiv \phi_i^2$ direction is 
$\sum_{j}|h_{ij}|^2|\phi_i|^6/\mpl^2$, where
$h_{ij} = \mpl m_{ij}/v^2$, and $v=175\gev$.  Given our
expectations for neutrino masses and mixing angles from the
MSW solution to the solar neutrino problem together with the
see-saw mechanism, the $L_e H_u$ direction is the most important,
with the off-diagonal coupling
$h_{e\mu}\sim (10^{18}\gev)(10^{-4}\ev)/(175\gev)^2\simeq 4$ being
dominant~\footnote{For the other $L_{\mu,\tau}H_u$ directions the couplings
are substantially larger, and therefore produce a much lower
final baryon to entropy ratio than the dominant $L_eH_u$ direction.}.
Fortunately this is in line with our naive expectations and
we will henceforth assume that the dominant $h$'s are $\CO(1)$.

In any case, after the inflationary phase has ended, the energy density
formerly in the $D$-terms is converted to $F$-term and kinetic energy
of the $S$ and $\psi_-$ fields, and thus the $\De m^2$ term can naturally
have the value $\rho_{\rm tot}/\mpl^2 \simeq H^2$.
For successful AD baryogenesis, the overall sign of the $H$-dependent
mass-squared term must be negative; we assume in the following
that the sign of the non-minimal K\"ahler terms are such that
this is the case. 

For $H^2\gg\mgrav^2$, 
the $-H^2|\ph|^2$ term completely dominates the other mass
terms in Eq.~(\ref{eq:flatV}), and $\vev{\ph}$
is determined by the balance between the negative $H$-dependent
(and thus time-dependent) effective mass and the leading
non-renormalizable term that lifts the flat direction
\beq
\vev{|\ph(t)|^2}
\simeq \left( {H(t)^2 \mpl^{2(n-3)}\over h^2} \right)^{1/(n-2)}.
\label{eq:phvev}
\eeq
The oscillations of $\ph$ around this expectation value are
critically damped for all values of $H>\mgrav$, so $\vev{|\ph|^2}$ accurately
tracks Eq.~(\ref{eq:phvev}).  In addition, the existence of 
this minimum for $|\ph|^2$ eliminates any initial value for $|\ph|$ which
may have been generated during inflation by, \eg, quantum fluctuations.
Thus it is not enough to consider quantum fluctuations alone as setting
non-zero values for $|\ph|$, since any such values will be washed out
by the post-inflationary potential in Eq.~(\ref{eq:phvev}).

This statement is also true of any initial value for $|\ph|$ set by the
inflationary $D$-term. For example, one could imagine the following seemingly
attractive scenario (which is a concrete realization of the suggestion
in~\cite{cg}): If the fields of the AD condensate are originally 
uncharged under U(1)$_{FI}$ but pick up a small charge through kinetic
mixing, then according to Eqs.~(\ref{eq:Lexpectation})--(\ref{eq:VwithKM}) 
the AD condensate will be
displaced from the origin at the end of inflation. However as soon as inflation
ends and the AD condensate gains an effective mass term $m_\ph^2\sim \pm H^2$,
its equation of motion will drive $\ph$ to its new minimum, which is
either at the origin (if $m^2_\ph>0$) or given by Eq.~(\ref{eq:phvev}) (if
$m_\ph^2<0$). Since the equation of motion is critically damped, the field 
will go to its new minimum in roughly a Hubble time. Therefore by the time of
baryogenesis, $|\ph|$ will have lost all information about its value
during inflation.

The same, however, is not true for the phase of the AD field. Because in
$D$-term inflationary models there are no $H$-dependent $A$-terms in the
post-inflationary potential, the equation of motion for $\arg(\ph)$ is
over-damped and $\arg(\ph)$ does not change until $H\simeq\mgrav$. (The
only term in the potential for $\arg(\ph)$ are the usual supergravity
$A$-terms. These lead to terms in the equation of motion for $\arg(\ph)$ 
which are small compared to the friction term 
until $H\simeq\mgrav$.) Therefore the value of the phase of $\ph$ and its
correlation length as set during the inflationary epoch are the values
relevant for AD baryogenesis, quite unlike the case of the magnitude of
$\ph$. In particular, if we are to get a net $B$ or $L$, the correlation
length for $\arg(\ph)$ must be greater than the current Hubble radius of 
the universe.

What sets the correlation length for the phase of $\ph$? There are two
cases to consider: one in which $|\ph|$ is massive and thus $\vev{\ph}=0$,
and the other in which $|\ph|$ is massless and thus $\vev{\ph}$ can be
large. In the first case, which for instance occurs when the U(1)$_{FI}$
charges for $\ph$ are positive and $\CO(1)$, the correlation length of
quantum de~Sitter fluctuations is~\cite{desitter}
\beq
\ell=H_{\rm infl}^{-1}\exp\left(\frac{3H_{\rm infl}^2}{2m_\ph^2}\right)
\label{eq:desitter}
\eeq
where $m_\ph^2$ is the mass-squared of $\ph$. Because $\vev{\ph}$ is zero 
classically the phase is ill-defined at the classical level. Therefore
any quantum fluctuations of $\ph$ pick up random phases with correlation
lengths given above. For $H_{\rm infl}=g^2\xi^4/6\mpl^2$ and $m_\ph^2\simeq
g^2\xi^2$, the exponent in the correlation length is much smaller than
the factor of 55 necessary to fit the entire observable universe
into one correlation volume. {\it AD baryogenesis cannot possibly
work in this case!}\/ Thus we derive the result that the AD field
must be a flat direction not only of the MSSM $D$-terms but also of the 
U(1)$_{FI}$ $D$-terms even though the U(1)$_{FI}$ interactions have been
integrated out at energies far above those at which baryogenesis is occurring.
Notice that a special example of this occurs for AD composites in which the
individual fields have bare $q=0$, but pick up non-zero $q$ via kinetic mixing.
Because the AD composite is by definition flat under hypercharge, it will
also be flat under U(1)$_{FI}$ even if the component fields are now charged.

For the second, massless case (\ie, $m^2_\ph\simeq\mgrav\ll H$) the correlation
length for de~Sitter fluctuations is enormous, much larger than the
observable universe. Thus any phase generated for $\ph$ will be coherent
and can lead to AD baryogenesis. This phase however is completely random
and in particular has no reason to be at the minimum selected by the the 
supergravity $A$-terms. And since the equations of motion for $\arg(\ph)$
are over-damped until $H\simeq\mgrav$, the value set by de~Sitter fluctuations
is the appropriate initial value for AD baryogenesis.
This is to be contrasted with the situation
for the magnitude of $\ph$, for which the value
resulting from de~Sitter fluctuations is washed out by the
$-H^2|\ph|^2$ terms that arise after inflation ends. 

However, there are in principle other processes which could destroy the 
coherence of the AD condensate in the post-inflationary period, 
in particular scattering off a thermal bath of
$\psi_-$ or $S$ decay products. If these processes are efficient,
then coherence length for the AD field will only be $\ell\simeq T^{-1}$,
much smaller than the radius of the observable universe, and so no net
baryon number will be created. In the Appendix we derive the result that
such scattering processes are reasonably efficient for AD flat directions
lifted at $n=4$ (\eg, the $LH_u$ direction), 
but completely inefficient for $n>4$, assuming
$\psi_-$ decays before the epoch of baryogenesis. Thus we view the
$n=4$ AD directions as disfavored unless one can arrange $\Ga_{\psi_-}
<\mgrav$. Such a suppression of the $\psi_-$ width is possible, but as 
argued in Section~\ref{sec:reheat}\ this requires that none of the MSSM fields
have bare U(1)$_{FI}$ charges and that the kinetic mixing be rather small
($\chi<10^{-6}$).

Now we turn to the actual baryon asymmetry production mechanism.
After the end of inflation, 
the system evolves, with the total energy density, Hubble constant, and
value of $\vev{|\ph(t)|^2}$, gradually decreasing until the supergravity
terms in the potential Eq.~(\ref{eq:flatV}) for the AD direction
become comparable to the $-H^2|\ph|^2$ term.  At this time
($H\simeq\mgrav$) the AD field starts oscillating around zero.
Most importantly, at this stage {\it all}\/ the terms in the potential 
for $\ph$
are of comparable magnitude, both baryon (or lepton) number conserving
and violating.  Also the phase of the $A$ terms together with
the {\it a priori}\/ random initial phase of the $\ph$ direction
provides significant CP-violating phases in the evolution of the
$\ph$ direction as it oscillates around zero. 
In these circumstances the AD condensate acquires a baryon
or lepton number per condensate particle of magnitude $\ep\sim\CO(1)$.
The resulting baryon to entropy ratio is given by~\cite{drt}
\beq
\frac{n_b}{s} \simeq \frac{n_b}{n_\ph} \frac{T^{(2)}_R}{m_\ph}
\frac{\rho_{\rm AD}}{\rho_{\rm tot}},
\label{eq:btos}
\eeq
where the first factor gives the baryon (or lepton) number per particle
in the AD condensate, and the second and third factors arise from converting
energy densities to entropy and number densities via
$\rho_{\rm AD}=m_\ph n_\ph$ and $\rho_{\rm tot} \simeq T_R s$.
As described in Section~\ref{sec:reheat}, the dilution in the baryon to 
entropy ratio due to a second stage of reheating is (up to a factor of
$\CO(1)$) taken account of by using the second stage reheat
temperature $T_R^{(2)}$ in the expression Eq.~(\ref{eq:btos}).
Substituting $\rho_{\rm AD}/\rho_{\rm tot}$, and taking
$m_\ph\sim \mgrav$, gives
\beq
\frac{n_b}{s} \simeq \ep \frac{T^{(2)}_R}{\mgrav}
\left(\frac{\mgrav^2}{h^2 \mpl^2 (n-2)}\right)^{1/(n-2)}.
\label{eq:btos2}
\eeq
Requiring this to reproduce the observed baryon to entropy ratio,
$n_b/s \simeq (3-10)\times 10^{-11}$,
leads to a correlation between the order, $n$, at which the flat
direction is lifted, and  $T^{(2)}_R$ (equivalently the decay width
of $S$).\footnote{In principle, there is another constraint
on our baryogenesis models that should be considered.  The
superpotential operator $h(LH_u)^2/\mpl$ is the MSSM operator
that gives rise to Majorana masses for the left handed neutrinos.
This operator, together with sphaleron processes, can destroy
any asymmetries in both $B$ and $L$, if the lepton-number
violating interactions it mediates are in equilibrium after
AD baryogenesis and reheat~\protect\cite{nelsonbarr}.  The resulting
constraint on the post-baryogenesis reheat temperature is very weak
$T_R \lsim 10^{16}\gev$ for an MSW-inspired spectrum of neutrino
masses and mixings.}

Given this, and further assuming that there is no {\it later}\/ dilution
of the baryon number (we will consider this possibility below), we now
summarize the resulting baryon to entropy ratio as a function of the order
at which the AD potential is lifted:
\begin{itemize}
\item
For $n=4$ we have already argued that reheating through $\psi_-$ decays
tends to decohere the AD condensate and therefore if $\psi_-$ decays
before the epoch of baryogenesis (at $H\simeq\mgrav$) no net baryon 
number is created. However, if $\psi_-$ can be made sufficiently long-lived
(\ie, $\Ga_\psi<\mgrav$), then baryogenesis takes place in a cold universe
(note that this requires the kinetic mixing $\chi<10^{-6}$).
In order to generate a final $n_b/s\simeq (3-10)\times10^{-11}$ then requires
$T_R^{(2)}\simeq10^7-10^8\gev$. This is
near the upper limit possible given the gravitino
constraint, and nicely fits in with the reheat temperature
expected from $S$-decay, Eq.~(\ref{eq:sreheat}), if we take
the leading allowed coupling of $S$ to MSSM fields to be
of the form $W=S\phi^4/\mpl^2$.

\item 
For $n=5$, $\psi_-$ decays can occur either before or after baryogenesis 
without adversely affecting the final baryon to entropy ratio. To derive
the measured $n_b/s$ along an $n=5$ direction requires
$T_R^{(2)} \simeq 10^3\gev$. Notice that this is quite small compared to 
some of our estimates of typical $S$ reheat temperatures. Thus baryogenesis
along an $n=5$ (or $n>5$) flat direction runs the risk of {\it over}producing
baryons which must then be diluted by an additional dump of entropy at
late times. In particular if $T_R^{(2)}\simeq10^8\gev$, then the baryons
are overproduced by a factor of $\CO(10^5)$ without a later dilution.
On the other hand, if the leading coupling of $S$ to MSSM fields occurs
at $r=6$, then this scenario leads to $n_b/s$ of the correct magnitude.

\item 
For $n=6$, the necessary reheat temperature is
$T_R^{(2)} \simeq 1\gev$, though larger temperatures again
overproduce baryons. For our naive estimate $T_R^{(2)}$ this overproduces
baryons by $\CO(10^8)$; although, again, $S$ decays might be greatly 
suppressed.

\item
For $n=7$, $T_R^{(2)}$ approaches the BBN bound of $6\mev$, and so for
$n>7$ it is always necessary to have a late time entropy dump in order
to avoid large $n_b/s$.
\end{itemize}
Notice that for the $n=4$ and $n=5$ directions, the final reheat temperature 
is above the weak scale and so sphaleron processes are active. Therefore
in these two cases, baryon number can be created through leptogenesis
followed by sphaleron-induced conversion to baryons as long as the AD 
condensate had net $B-L$.

Finally one should always bear in mind that supersymmetry models, and in 
particular supersymmmetry-breaking models, 
generically bring with them other moduli 
from either the hidden sector or from the string sector which have naturally
late decays~\cite{moduli}, spoiling the success of Big-Bang nucleosynthesis. 
If the necessary dilution of these moduli occurs after baryogenesis, 
the large increase in 
the entropy will also dilute the baryon number by many orders of
magnitude. Therefore it is quite possible that baryogenesis must be
over-efficient in the early universe in order to produce the current 
$n_b/s$ ratio of roughly $10^{-10}$ as a ``post-dilution'' value.
It is comforting to see that, except along the 
$n=4$ direction, it is possible to greatly overproduce baryons in the 
AD scenarios we have investigated. This is an option that other
forms of baryogenesis, \eg\ electroweak baryogenesis, do not provide.

\section{Conclusions}\label{sec:conc}

We have demonstrated in this paper several important results about 
$D$-term inflation. First, we have found that inflation ends in these
models at field values of $S$ far above its critical value, and that in order
to have sufficient e-folds of inflation, the initial value must be 
quite large and the potential unusually flat. In particular we found that
the potential for $S$ must be flat out to dimension-10 terms in the 
superpotential, a requirement that seems to point to the existence of 
$R$-symmetries which protect the flatness of the $S$ potential.

Second, we have found that reheating in these models is generically immediate
because of the $\CO(1)$ couplings of the $\psi_-$ field to light matter. The
corresponding first-stage reheat temperature is about $10^{15}\gev$. 
Even if we tried to further push down
that reheat temperature by constraining the couplings of $\psi_-$
to light matter, we found that it would be an unnatural tuning
of the model for its temperature to fall below about $10^{10}\gev$. 
Nevertheless, the $D$-term models very nicely overcome the gravitino 
constraints since they possess a second condensate (the $S$-field) whose
suppressed decay width naturally leads to a reheat temperature 
$T_R<10^9\gev$.

Finally, we have shown that Affleck-Dine baryogenesis can proceed in models of
$D$-term inflation in much the same way in which one expected it in 
models of $F$-term inflation. That is, even though there are not $F$-terms
present during inflation, they do appear at the end of inflation in the form
of either oscillating $\psi_-$ and $S$ fields, or in the finite energy 
densities present in the plasma after their decays. This energy density can,
given a needed sign in the K\"ahler potential, produce negative effective
squared-masses for the AD condensate, pushing its modulus 
away from its true minimum at the origin until $H\simeq\mw$. Meanwhile
the phase of the AD condensate has been set by de~Sitter fluctuations which
are inflated to superhorizon-size correlation lengths. The mismatch between
this random phase and the phase preferred by the low-temperature scalar 
potential provides the $CP$ violation necessary for baryogenesis.

We find that the high-temperature
environment produced by the $\psi_-$ decays can decohere the AD condensate
if its potential is only lifted at $n=4$ in the superpotential; however,
AD flat directions lifted at $n>4$ stay coherent over super-horizon volumes
and thus produce substantial baryon number. 

Thus $D$-term inflationary models
can accomodate successful phenomenology, including very flat slow-roll
potentials mandated by symmetries, reheating consistent with gravitino and 
nucleosynthesis constraints, and an efficient and appealing baryogenesis
mechanism.


\section*{Acknowledgments}
We are grateful to M.~Alford, H.~Murayama, L.~Randall, M.~Strassler, and
particularly K.S.~Babu for discussions.


\section*{Appendix}
In this Appendix we will show that the thermal bath of MSSM particles
created by decay of
the $\psi_-$ field can decohere the phase of the AD condensate when it is
lifted at $n=4$ order, but not if lifted at $n\geq5$.

Consider first scattering of thermal MSSM fields off of the AD condensate via
gauge interactions. Any gauge boson which couples to the AD field will
get a mass $g\vev{\ph}$ so that the thermally averaged scattering rate
per AD mode is approximately
\beq
\Ga_{\rm scatt}= n_\phi\vev{v\sigma}\simeq
\frac{g^4 T^2}{16\pi(T^2-g^2\langle\ph\rangle^2)^2}T^3.
\eeq
where we have assumed that the particles in the thermal bath are relativistic.
(This should always be the case since, \eg, the selectron is always light
compared to $T$ or $\vev{\ph}$ and so is copiously produced in the thermal 
bath.)

For $n=4$ flat directions, $g\vev{\ph}\sim T\sim \sqrt{H\mpl}$ 
for $\psi_-$ decaying at $H\simeq\Ga_{\psi_-}$ 
(we assume $h\sim\CO(1)$ throughout this appendix). 
If $\psi_-$ decays immediately after
the end of inflation then $\Ga_{\rm scatt}\simeq T/16\pi\simeq\xi/16\pi
> H\simeq
\xi^2/\mpl$ and thus there are many scattering per Hubble time per AD mode
and the AD condensate is evaporated into an incoherent collection of individual
particles. If the $\psi_-$ decays late, but prior to $H\simeq\mgrav$, the
scattering rate is even larger compared to $H$ and so again the AD condensate
is evaporated. Of course, if $\psi_-$ decays after $H\simeq\mgrav$, that is,
after baryogenesis, then there is no difficulty getting the currently
observed baryon asymmetry.

For $n=5$ flat directions, $g\vev{\ph}\sim (H\mpl^2)^{1/3}\gg T\simeq
\sqrt{H\mpl}$
so that $\Ga_{\rm scatt}\simeq T^5/16\pi\ph^4$. It is easy to show that for
$\psi_-$ decays at any time after the end of inflation, $\Ga_{\rm scatt}\ll
H$ and so the AD condensate is {\it not}\/ evaporated by gauge interactions. 
This result holds also for all flat directions with $n\geq6$.

There is one other scattering process which could evaporate the AD condensate
coming from superpotential couplings: ${\CL}=\la^2 |\phi|^2 |\varphi|^2$
where $\la$ is a usual Yukawa coupling. It is useful to consider the
two limits in which the MSSM particles in the thermal bath are relativistic
or non-relativistic. These correspond respectively to small $\la\ll1$  
(typical of the first two generations of MSSM
squarks and sleptons) and to large $\la\sim1$ 
(typical of the third generation) since we expect that particles in the
thermal bath to acquire masses through the vev of $\ph$ which are of order 
$\la\vev{\ph}$.

Consider first the case of a non-relativistic thermal bath.
The non-relativistic scattering rate is calculated using 
$n_\phi=\rho_\phi/m_\phi$, relative velocity $v_\phi\simeq
\sqrt{T/m_\phi}$ and $\sigma\simeq\la^4/16\pi m^2_\phi$. 
Specifying an $n=5$ flat direction and
using the expression for $\vev{\ph}$ in Eq.~(\ref{eq:phvev}), one finds
\beq
\Ga_{\rm scatt}\simeq\la^{1/2}H(H/\mpl)^{1/12}/16\pi
\eeq
(again we are 
calculating at the point of reheat where $T\simeq\sqrt{H\mpl}$). Thus
$\Ga_{\rm scatt}< H$ and the AD condensate is not evaporated; 
this is even more
true as the reheat period approaches (from above) $H\simeq\mgrav$. Thus
the non-relativistic component of the thermal bath does not scatter
off the AD condensate with sufficient strength to decohere the phase of the
condensate. This argument generalizes for all $n\geq5$. 

For a relativistic thermal bath, the ``worst case'' occurs when 
$\psi_-$ decays right before baryogenesis (so $H\simeq\mgrav$). Then
\beq
\Ga_{\rm scatt}\simeq\frac{\la^4}{16\pi T^2}T^3\simeq
\frac{\la^4}{16\pi}\sqrt{\mgrav\mpl}
\eeq
which is less than $H$ as long as
$\la\lsim 10^{-2}$. (Note that this does not depend on the order at
which the AD flat direction is lifted.) But this is essentially the same
condition for the $\phi$ fields to be relativistic. In particular, for
immediate decay of $\psi_-$, one has $T>\la\vev{\ph}$ precisely for
$\la\lsim10^{-2}$. Therefore we conclude that the relativistic component
of the thermal bath also does not efficiently decohere the phase of an
AD condensate lifted at order $n\geq5$.


\begin{thebibliography}{99}

\bibitem{slowroll}
	A.~Albrecht and P.~Steinhardt, \PRL{48}{82}{1220};\\
	A.~Linde, \PLB{108}{82}{389}.

\bibitem{earlysusy}
	M.~Dine, W.~Fischler and D.~Nemeschansky,\PLB{136}{84}{169};\\
	G.~Coughlan, \etal, \PLB{140}{84}{44}.

\bibitem{copeland}
	E.~Copeland, \etal, \PRD{49}{94}{6410}.

\bibitem{drt}
	M.~Dine, L.~Randall and S.~Thomas, \NPB{458}{96}{291}.

\bibitem{early}
	J.~Casas and C.~Munoz, \PLB{216}{89}{37};\\
	J.~Casas, \etal, \NPB{328}{89}{272};\\
	E.~Stewart, \PRD{51}{95}{6847}.

\bibitem{ad}
	I.~Affleck and M.~Dine, \NPB{249}{85}{361}.

\bibitem{hitoshi}
	H.~Murayama and T.~Yanagida, \PLB{322}{94}{349}.

\bibitem{halyobd}
	E.~Halyo, \PLB{387}{96}{43};\\
	P.~Binetruy and G.~Dvali, \PLB{388}{96}{241}.

\bibitem{jeannerot}
	R.~Jeannerot, \PRD{56}{97}{6205}.

\bibitem{lyth}
	D.~Lyth, {\tt [hep-ph/9710347]}.

\bibitem{hybrid}
	A.~Linde, \PLB{259}{91}{38}; \PRD{49}{94}{748};\\
	A.~Liddle and D.~Lyth, \PRD{52}{95}{6789}.

\bibitem{planck}
	M.~Kamionkowski and J.~March-Russell, \PLB{282}{92}{137};\\
	R.~Holman, \etal, \PLB{282}{92}{132}.

\bibitem{DGT}
	L.~Krauss and F.~Wilczek, \PRL{62}{89}{1221};\\
	M.~Alford, J.~March-Russell and F.~Wilczek, \NPB{337}{90}{695};\\
	J.~Preskill and L.~Krauss, \NPB{341}{90}{50}.

\bibitem{open}
	J.~Gott, Nature, {\bf 295} (1982) 304;\\
	M.~Bucher, A.~Goldhaber and N.~Turok, \PRD{52}{95}{3314};\\
	K.~Yamamoto, M.~Sasaki and T.~Tanaka, Ap.~J. {\bf 455} (1995) 412;\\
	A.~Linde, \PLB{351}{95}{99};\\
	J.~Garcia-Bellido and A.~Linde, \PRD{55}{97}{7480}.

\bibitem{fiterm}
	M.~Dine, N.~Seiberg and E.~Witten, \NPB{289}{87}{589};\\
	J.~Atick, L.~Dixon and A.~Sen, \NPB{292}{87}{109}.

\bibitem{LR}
	D.~Lyth and A.~Riotto, {\tt [hep-ph/9707273]}.

\bibitem{matsuda}
	T.~Matsuda, {\tt [hep-ph/9705448]}.

\bibitem{moroi}
	J.~Ellis, A.~Linde and D.~Nanopoulos, \PLB{118}{82}{59};\\
	Also see, for example;\\
	T.~Moroi, Ph.D. Thesis, Tohoku University (1995),
	{\tt [hep-ph/9503210]}.

\bibitem{bbn}
	M.~Reno and D.~Seckel, \PRD{37}{88}{3441};\\
	G.~Lazarides, \etal, \NPB{346}{90}{193}.

\bibitem{param}
	I.~Zlatev, G.~Huey and P.~Steinhardt, {\tt [astro-ph/9709006]}.

\bibitem{holdom}
	B.~Holdom, \PLB{166}{86}{196}; \\
	F.~del~Aguila, G.D. Coughlan and M. Quiros, \NPB{307}{88}{633};\\
	K.~Babu, C.~Kolda and J.~March-Russell, \PRD{54}{96}{4635}.

\bibitem{dkmr}
	K.~Dienes, C.~Kolda and J.~March-Russell, \NPB{492}{97}{104}.

\bibitem{dflat}
	T.~Gherghetta, C.~Kolda and S.~Martin, \NPB{468}{96}{37}.

\bibitem{cg}
	J.~Casas and G.~Gelmini, \PLB{410}{97}{36}.

\bibitem{desitter}
	T.~Bunch and P.~Davies, Proc.\ R.\ Soc.\ London, {\bf A360}
	(1978) 117.

\bibitem{nelsonbarr}
	A.~Nelson and S.~Barr, \PLB{246}{90}{141}.

\bibitem{moduli}
	See for example:\\
	G.~Coughlan, \etal, \PLB{131}{83}{59};\\
	T.~Banks, D.~Kaplan and A.~Nelson, \PRD{49}{94}{779};\\
	L.~Randall and S.~Thomas, \NPB{449}{95}{229}.

\end{thebibliography}
\end{document}